\documentclass[a4paper,fleqn,usenatbib]{mnras}

\usepackage{mathptmx}
\usepackage{gensymb}
\usepackage{comment}
\usepackage{bm}

\usepackage[T1]{fontenc}
\usepackage{ae,aecompl}
\usepackage[section]{placeins}

\usepackage{graphicx}	
\usepackage{amsmath}	
\usepackage{amssymb}	
\usepackage{multicol}        
\usepackage{pdflscape}
\usepackage{float}
\usepackage{longtable}
\usepackage{breqn}

\newcommand{\msun}{M$_{\odot}$}

\newcommand{\e}[1]{$\times10^{#1}$}
\newcommand{\invcmcube}{cm$^{-3}$}
\newcommand{\kmpers}{km s$^{-1}$}
\newcommand{\hydra}{V Hya}

\title[Unseen companions of V Hya]{Unseen companions of V Hya inferred from periodic ejections}
\author[J. M. Salas et. al.]{
	Jesus M. Salas,$^{1}$\thanks{E-mail: jesusms@astro.ucla.edu}
		Smadar Naoz$^{1,2}$,
	 Mark R. Morris$^{1}$ and
     Alexander P. Stephan$^{1,2}$	
	\\
	$^{1}$Dept. of Physics \& Astronomy, University of California, Los Angeles, CA, 90095, USA	
     \\ 
    $^{2}$Mani L. Bhaumik Institute for Theoretical Physics, Department of Physics and Astronomy, \\
University of California, Los Angeles, CA 90095, USA
}

\begin{document}
	\label{firstpage}
	\pagerange{\pageref{firstpage}--\pageref{lastpage}}
	\maketitle
	
\begin{abstract}
A recent study using \textit{Hubble Space Telescope} observations found periodic, high-speed, collimated ejections (or ``bullets") from the star V Hya. The authors of that study proposed a model associating these bullets with the periastron passage of an unseen, substellar companion in an eccentric orbit and with an orbital period of $\sim8$ yrs. Here we propose that V Hya is part of a triple system, with a substellar companion having an orbital period of $\sim8$ yrs, and a tertiary object on a much wider orbit. In this model, the more distant object causes high-eccentricity excitations on the substellar companion's orbit via the Eccentric Kozai-Lidov mechanism. These eccentricities can reach such high values that they lead to Roche-lobe crossing, producing the observed bullet ejections via a strongly enhanced accretion episode. For example, we find that a ballistic bullet ejection mechanism can be produced by a brown-dwarf-mass companion, while magnetically driven outflows are consistent with a Jovian-mass companion.  Finally, we suggest that the distant companion may reside at few a hundred AU on an eccentric orbit.  
\end{abstract}


\section{Introduction}
Observations from the $Hubble$ $Space$ $Telescope$ (HST) over the past two decades have revealed an enormous complexity and diversity of structure in planetary nebulae \citep[PNe;][]{Balick2002ARA&A..40..439B, Sahai2011AJ....141..134S}. HST surveys have revealed that more than half of PNe are bipolar or multipolar, whereas mass loss during the AGB phase is mostly spherical.  This led \cite{Sahai1998} to propose that high-speed, collimated (jet-like) outflows during the late AGB phase that can change their orientation could be the reason behind the asymmetric morphology of PNe. These jets could be driven by interactions with a binary companion \citep{Morris1987}; however, direct evidence supporting this idea has been lacking.

The carbon star \hydra\ is one example where there exists evidence for high-speed, collimated outflows 
\citep{Evans1991,Knapp1997,Sahai1998,Sahai2003,Hirano2004,Sahai2009}. A recent study by \cite{Sahai2016} presents new HST observations that span more than a decade and provide, with an unprecedented and detailed view, the extended history and characteristics of the bullet-like ejections from \hydra. Their data show that these high-speed ($\sim$$200-250$ \kmpers) bullets are ejected once every $\sim8.5$ yrs, and that the axis of ejection flip-flops around a roughly eastern direction, both in and perpendicular to the sky plane. 

To account for this phenomenon, \cite{Sahai2016} proposed a model in which the bullets are associated with the periastron passage of a binary companion in an eccentric orbit with an orbital period of $\sim$8.5 years. The bullets are likely ejected from an accretion disk formed around the companion that results from the gravitational capture of matter levitated into the primary's wind-formation zone, and perhaps directly from the primary's pulsating atmosphere. However, this hypothesis faces the difficulty that tidal forces between binary companions tend to shrink and circularize their orbits, or even cause mergers.  To overcome this problem, we here propose a more elaborated model in which \hydra\ is part of a triple system. In such a system, a relatively distant third object can impose an eccentric orbit on the inner companion, and even lead to Roche-limit crossing, thus allowing the inner companion to accrete and eject mass. 

Studies of stellar populations have shown that multiple star systems are very common, with $\sim50\%$ of Sun-like stars having binary companions, and even higher
fractions ($\sim70\%$) are found for higher-mass stars (e.g., \citealt{Raghavan+2010}). Moreover, it seems that many of these binaries are in triples or higher multiples. For example, \citet{Tokovinin1997} showed that $\sim40\%$ of short-period binary stars with a low-mass ($\sim$$0.5–-1.5$\msun) primary have at least one additional companion. Furthermore, among contact binaries it seems that about $42\%$ are in a triple configuration (e.g., \citealt{Pribulla+2006}). These and many other observational endeavors have revealed that triple star systems are common (e.g., \citealt{Tokovinin1997b,Tokovinin+2006,Eggleton2007,Griffin2012}. See also \citealt{Tokovinin2008,Tokovinin2014a,Tokovinin2014b}).

Dynamical stability considerations dictate that triple systems must be hierarchical in scale, in which the (``inner'') binary is orbited by a third body on a much wider (``outer'') orbit. In this setup, the inner binary undergoes large-amplitude eccentricity and inclination oscillations due to the ``Eccentric Kozai-Lidov" (EKL) mechanism (\citealt{Kozai1962AJ.....67..591K,Lidov1962P&SS....9..719L}.
For a review, see \citealt{Naoz2016}). These eccentricity excitations can drive the inner binary to have very small pericenter distances and even to merge \citep[e.g.,][]{Naoz2014,Prodan2015ApJ...799..118P,Naoz2016,Stephan2016,Stephan2017,Stephan2018AJ....156..128S}.

The star \hydra\ is currently in its AGB stage with a mass of $\sim$$1-2$ \msun\ and a radius of $\sim2$ AU (see \citealt{Zhao-Geisler2012}). The rapid evolution of an AGB star in a triple system can play a major role in the dynamical evolution of such a system. (e.g, \citealt{Perets2012,Shappee2013,Toonen2016ComAC...3....6T,Michaely2016,Naoz+2016,Stephan2016,Stephan2017,Stephan2018AJ....156..128S}). For example, as the AGB star loses mass, it can reduce the separation ratio between the inner and outer orbits, thus, re-triggering EKL eccentricity excitations \citep{Shappee2013}. In addition, as the star expands, tidal forces become more efficient since tides are highly sensitive to the stellar radius.

In particular for the \hydra\ system, EKL combined with post-main-sequence stellar evolution can drive the inner binary to very high eccentricities and even cause it to undergo Roche-lobe crossing, inducing the secondary object to accrete material from its companion, and perhaps eject some of this material from an accretion disk \citep{Sahai2016}. The mechanism we introduce here thus could explain the observed ejections from \hydra.

This paper is organized as follows: in Section \ref{sec:numerical_methods} we describe the code and numerical setup. Our results are shown in Section \ref{sec:results}. A discussion of the implications of our model in Section \ref{sec:implications}, and conclude in Section \ref{sec:discussion}.

\section{Numerical Methods}\label{sec:numerical_methods}
\subsection{Stellar evolution and three-body dynamics}
We solve the secular equations for a hierarchical triple system up to the octupole level of approximation (as described in \citealt{Naoz2013a,Naoz2016}), including general relativistic effects for both the inner and outer orbit \citep{Naoz2013b} and static tides for both members of the stellar binary (following \citealt{Hut1980,Eggleton1998}, see \citealt{Naoz2016} for the complete set of equations). We also include the effects of stellar evolution on stellar radii and masses, following the stellar evolution code \texttt{SSE} by \cite{Hurley2000}. The interaction between the EKL mechanism and post-main-sequence stellar evolution has been demonstrated to play an important role in three-body dynamical evolution (see \citealt{Perets2012,Shappee2013,Toonen2016ComAC...3....6T,Michaely2016,Naoz+2016,Stephan2016,Stephan2017,Stephan2018AJ....156..128S}).

\subsection{Numerical setup}
We divide the parameter space into a grid in which we choose among a set of initial values for the masses of the three bodies ($M_{\hydra}$, $m_1$, $m_2$), the semi-major axes of the inner and outer orbits ($a_1$, $a_2$), the eccentricities of the inner and outer orbits ($e_1$, $e_2$), and the inclination ($i$) between the two orbits\footnote{Throughout this paper, we use the subscripts $1$ and $2$ to indicate the values for the inner and outer orbits of the system, respectively.  For the mass parameter, the subscripts indicate \hydra\ ($M_{\hydra}$), the inner companion ($m_1$) and the outer body ($m_2$)}. 

Based on SSE modeling, the Zero-Age Main Sequence (ZAMS) mass of \hydra\ ($M_{\hydra}$) was set to $2.2$ \msun, appropriate for a carbon star. Each system is integrated for $1.2$ Gyrs unless a stopping condition is fulfilled. To allow for comparison with the observed system, we focus on the Late-AGB phase (L-AGB, i.e., $1.143$ to $1.146$ Gyr for the chosen mass) and determine whether the periastron of the inner orbit  reaches the primary's Roche limit without merging. The Roche limit of V Hya is defined as \citep[e.g.,][]{Pacynski1971ARA&A...9..183P,Matsumura2010ApJ...725.1995M,Naoz2016}:
\begin{equation}\label{eq:roche}
RL_{\hydra} =  q R_{\hydra} \left( \frac{M_{\hydra} + m_1}{M_{\hydra}}  \right)^{1/3}
\end{equation}  
where $q$ is a numerical factor of order unity. However, the radius of an AGB star is not well defined; its stellar envelope can extend to large distances, perhaps filling its own Roche limit. Thus, it is reasonable to assume that mass accretion can occur when the inner companion reaches the primary's Roche limit. We note that the parameter $q$ is rather uncertain, in particular for a bound eccentric case. Numerical simulations have suggested that this parameter can be about $2.7$ (e.g., \citealt{Guillochon2011ApJ...732...74G,Liu2013ApJ...762...37L}), and that value may be a lower limit. Other studies sometimes find and adopt a  smaller value ($q\sim 1.4-1.6$, e.g., \citealt{Pacynski1971ARA&A...9..183P}). Thus, here, we explore two limiting cases, one for which $q=1.66$, as was adopted in \cite{Naoz2012ApJ...754L..36N}, and another for which $q=2.7$, adopted in \cite{Petrovich2015ApJ...799...27P}. 
These two limiting cases represent two different physical pictures: interactions with an extended envelope ($q=2.7$) and a contained envelope ($q=1.66$).
The integration was stopped when the inner orbit pericenter $R_{p,1}$ reaches $80\%$ of \hydra's radius, $R_{\hydra}$, i.e., $R_{p,1} \le 0.8 R_{\hydra}$. 

We investigated a discrete range of initial values for $m_1$; a Neptune-sized planet ($5$\e{-5} \msun), a Jupiter-sized planet ($10^{-3}$ \msun), a brown dwarf ($0.01$ \msun), and a range of subsolar stellar companions ($0.1$, $0.3$, $0.6$, $0.9$ \msun). According to \cite{Sahai2016}, the inner companion's mass should be sub-solar, thus we do not consider larger mass companions in our investigation. 

The semi-major axis of the inner orbit ($a_1$) was set so that its period is $8.5$ years. We also note that it is unlikely that a planet around \hydra\ would have a high initial eccentricity. On the other hand, a stellar companion could have had a high initial eccentricity, but tides would have circularized its orbit by the time the primary entered the AGB phase. Thus, for the eccentricity of the inner companion, we adopt for simplicity an initial, almost circular orbit ($e_1 = 0.1$). 

There is a degeneracy between $m_2$, $a_2$ and $e_2$ that comes from the Kozai timescale \citep[e.g.,][]{Naoz2016}:
\begin{equation}
t_{quad} \propto \frac{a_2^2 (1-e_2^2)^{3/2} \sqrt{M_{\hydra} + m_1}}{a_1^{3/2} m_2}.
\end{equation}
and therefore we can restrict these parameters to a narrow range of values because this time scale must be shorter than the lifetime of \hydra. However, if $m_2$ is larger than $m_1$, the system dynamics can be described well by a test particle approximation ($m_2$ $\gtrsim$ $7m_1$, \citealt{Teyssandier2013}). In this case, the inner orbit can reach extreme eccentricities in very short
timescales (e.g., \citealt{Li2014ApJ...785..116L, Li2014ApJ...791...86L}). Thus, we do not expect differences in the evolution of systems for which $m_2$ $\gtrsim7m_1$.  We chose a lower limit of $m_2$ to be $0.01$ \msun\ (a brown dwarf). As with $m_1$, we chose an upper limit of $m_2$ = $0.9$ \msun.

In the model we propose here, we assume that \hydra\ is part of a hierarchical triple system. This means that the value for $a_2$ must be much greater than $a_1$. Such a configuration allows us to use of the secular approximation equations \citep{Naoz2016}. Furthermore, $a_2$ needs to satisfy the following criterion for the secular approximation to be valid (e.g., \citealt{Lithwick2011ApJ...742...94L}):
\begin{equation}
\epsilon = \frac{a_1}{a_2} \frac{e_2}{1 - e_2^2} \  < 0.1,
\end{equation}
where $\epsilon$ is a measure of the relative strengths of the octupole and quadrupole effects on the orbital dynamics. Therefore, we test a range of  200 to 1000 AU in 200 AU increments, with eccentricity values ($e_2$) of $0.3$, $0.45$ and $0.6$. 

Finally, we test a wide range of mutual orbit inclinations ($i$ = $35^{\circ}$ to $175^{\circ}$ in $35^{\circ}$ increments). Table \ref{table:IC2} summarizes the parameters of our computations, which give a total of $2625$ cases that were generated as initial conditions.  
\begin{table}
\centering
	\caption{Grid of initial conditions for our 3-body model. A combination of all of these values gives a set of $2625$ initial conditions. Parameters marked with a (*) were set the same for all computations. The semi-major axis of the inner orbit for each $m_1$ was calculated via $a_1$ = ($P_1^2(M_{\hydra}+m_1))^{1/3}$.}
	\begin{tabular}{lc}
		Parameter & Initial values \\
		\hline
        $M_{\hydra}$* & 2.2 (in \msun) 
        \\
		$m_1$ & 5\e{-5}, 0.001, 0.01, 0.1, 0.3, 0.6, 0.9 (in \msun) 
        \\
		$m_2$ & 0.01, 0.1, 0.3, 0.6, 0.9 (in \msun) 
        \\
        $a_1$* & Set such that $P_1$ = 8.5 yrs ($\sim5-6$ AU)
        \\
		$a_2$ & 200, 400, 600, 800, 1000 (in AU) 
        \\
        $e_1$* & 0.1
        \\
        $e_2$ & 0.3, 0.45, 0.6
        \\
        $i$ & 35, 70 105, 140, 175 (in degrees)
	\end{tabular}
\label{table:IC2}
\end{table}

\section{Results}\label{sec:results}
In this Section we present the results from our $2625$ simulated triple systems. We note that less than $1\%$ of our simulated systems were inconclusive, and thus we ignore those systems in our analysis. 

We divide the simulations into ``survived" and ``merged" systems. We show examples in Figure \ref{fig:sur_merge}, which presents the time evolution of the inner orbit's semi-major axis (red), periastron distance (blue), and \hydra's stellar radius (purple) and Roche limit (green and cyan dashed lines). The Late AGB phase is shaded in purple, which lasts for $\sim3$ Myrs for the chosen initial value of $M_{\hydra}$ = $2.2$ \msun. 
\begin{itemize}
\item \textbf{Merged systems}: Here we include all systems in which the inner binary merged at any point of the evolution, which occurs in $\sim37\%$ of all simulated cases (991/2625). These merged systems can be divided into two groups: 
\begin{enumerate}
\item  \textbf{L-AGB mergers}: systems which merged during the L-AGB period (an example is shown in the middle panel of Figure \ref{fig:sur_merge}). These can happen via an EKL-induced high eccentricity (e.g., \citealt{Shappee2013,Naoz2014,Stephan2018AJ....156..128S}). Furthermore, because tides are highly sensitive to the stellar radius, the likelihood of a merger is increased at this stage of stellar evolution due to circularization and shrinking of the inner binary's orbit (e.g., \citealt{Naoz2016}). These systems comprise $\sim76\%$ of all mergers (752/991). 
\item \textbf{Pre-AGB mergers}: systems which merged in previous stages of the stellar evolution of \hydra\ (right panel of Figure \ref{fig:sur_merge}). These occur due to strong EKL effects due to stellar-mass inner and outer companions, as well as strong tidal interactions between the inner binary members. Such systems are likely to give rise to blue stragglers (e.g., \citealt{Perets2009ApJ...697.1048P,Naoz2014}). These systems comprise $\sim24\%$ of all mergers (752/991). 
\end{enumerate}
\item \textbf{Survived binary systems}: Here we include systems in which the inner companion survives the evolution of \hydra\ without merging. This occurs in $\sim62\%$ of all simulated cases (1616/2625). In these systems, the semi-major axis of the inner orbit increases substantially at $\sim1.146$ Gyrs, which is when the star sheds its outer layers and becomes a white dwarf. Most of these systems are ``no interaction'' systems, in which the inner companion survives the evolution of V Hya without interacting with the primary's Roche limit (an example of such a system in shown in the left panel of Figure \ref{fig:sur_merge}). In general, these are systems in which EKL effects were insufficient to induce high eccentricities due to, for example,  nearly coplanar orbits\footnote{Coplanarity does not guarantee a small eccentricity excitation, as was shown in \cite{Li2014ApJ...785..116L}} ($i$ = 35 or 175 degrees, such as the example in the left panel of Figure \ref{fig:sur_merge}), high $a_2$ values, or ratios $m_2/m_1$ $\approx$ $1$. However, there are also systems where EKL effects did increase the inner orbit's eccentricity, but not enough to make the inner orbit's periastron cross the primary's Roche limit (see Figure \ref{fig:correlations}). Our investigation focuses on the subset of surviving systems in which the inner companion's periastron reaches the primary's Roche limit during the L-AGB phase. These occur in $\sim8\%$ of all cases
(examples of these systems' orbital evolution are shown in Figure \ref{fig:grazing}), and can be further divided into two subcategories:
\begin{enumerate}
 \item \textbf{``Grazing" systems}: where the inner companion's orbit reaches a high eccentricity ($e_1$ > 0.1) and crosses the primary's Roche limit during its periastron passage (examples are shown in the left and middle panels of Figure \ref{fig:grazing}). This configuration is maintained for $\sim1$ Myr until the primary becomes a white dwarf, at which point the inner companion's orbit increases and exits the Roche limit. During this 1 Myr period, $M_{\hydra}$ $\sim1.5-1.7$ \msun, and $R_{\hydra}$ $\sim1.6$ AU. In this configuration, the inner companion could, in principle, accrete material during its periastron passage. Then \hydra-like ejections could be produced by a transient accretion disk. The number of systems in this category depends on $q$. For $q$ = $2.7$, this condition is satisfied in $\sim9\%$ of surviving systems (138/1616), while it is satisfied in only $\sim2\%$ of surviving systems (27/1616) for $q$ = 1.66.
 \item \textbf{``Temporary Close Binaries" (TCBs)}: systems where the inner companion's orbit circularizes during the Late-AGB phase and is engulfed by \hydra's Roche limit, but does not lead to a merger with the latter (an example is shown in the right panel of Figure \ref{fig:grazing}). We call these systems ``Temporary Close Binaries". This condition is only satisfied using a factor of $q$ = 2.7 in the definition of the Roche limit, and in $\sim4\%$ of surviving systems (60/1616). While these cases could not produce \hydra-like systems, they are likely to end up as common envelope configurations because of the drag encountered inside the Roche limit\footnote{Drag forces also affect the grazing systems we describe above. To determine whether the primary's extended envelope significantly affects the grazing companion's orbit is equivalent to asking whether the orbital average is valid in the secular approximation, compared to the orbital (or less than orbital) effects due to the drag force. In other words, if on an orbital timescale the companion's velocity is changing significantly, then the secular approximation is invalid. So the question boils down to which force dominates over the orbital timescale (the EKL produces forced eccentricity which remains constant over the entire $\sim1$ Myr in which the companion's periastron grazes the Roche limit of V Hya). Estimating the drag force as $F_{drag} \approx 0.5 \rho \pi R_1^2 v_1^2$, using $\rho=10^{-8}$ g cm$^-3$ \citep{Lagadec2005A&A...433..553L}, we a find very small change in velocity of the grazing companions (Jovians and brown dwarf-mass companions) due to the drag force (<1$\%$) over one orbit, the velocity will have been re-established by EKL again.}.
\end{enumerate}

\end{itemize}
 
We show in Figure \ref{fig:correlations} how the mass of the inner companion ($m_1$) affects the evolution of the system. In the left panel we illustrate the relationship between $D_{min}$ and eccentricity. We define $D_{min}$ as:
\begin{equation}\label{eq:dmin}
D_{min} = min[ (a_1(1-e_1)) - RL_{\hydra} ],
\end{equation}
i.e., the minimum distance between the periastron of the inner orbit and \hydra's Roche limit ($D_{min}$ = 0 indicates Roche limit crossing, using $q$ = 2.7). On the $x$-axis, we show the eccentricity at the time of $D_{min}$ ($t_{D_{min}}$, which occurs during the L-AGB phase). It is clear that most surviving stellar companions ($m_1$ > 0.1 \msun) circularize (reach final values of $e_1$ $\sim$ 0). This is because stellar companions have larger radii than lower-mass objects, and since tides are highly dependent on the radius, their orbits circularize relatively quickly. Because of the relatively larger mass of stellar objects, the EKL mechanism isn't as effective since the distant tertiary has lower or similar mass. 

Most companions with $m_1$ = 0.1 \msun\ merge during or before the L-AGB phase, as indicated in the right panel of Figure \ref{fig:grazing}.  Most of the TCB systems contain a stellar mass $m_1$ $>$ $0.1$ \msun\ companion, while brown dwarfs and planets ($m_1$ $<$ 0.1 \msun) produce mainly grazing systems. 

Most surviving non-stellar companions ($m_1$ < 0.1 \msun, i.e., Jovians and Brown dwarfs) follow a linear relation between $D_{min}$ and $e_{1}(t_{D_{min}})$. There is also a subtle mass dependence (see Figure \ref{fig:correlations}): lower-mass objects reach higher eccentricities. Using the upper limit value of $q$ = $2.7$, our simulations show that among all the objects that end up grazing the primary's Roche limit, there are more Neptune-mass objects than any other, followed by Jupiter-mass and brown dwarf companions. This is due to the fact that, for larger $m_2/m_1$ ratios, EKL effects approach the test particle approximation, in which the inner orbit achieves very high eccentricities. Moreover, tides become less effective for inner companions with small radii. Therefore, systems with Neptune-mass inner companions can graze V Hya's Roche limit with the highest eccentricities without being significantly affected by tidal forces.

\begin{figure}
	\includegraphics[width=1 \textwidth]{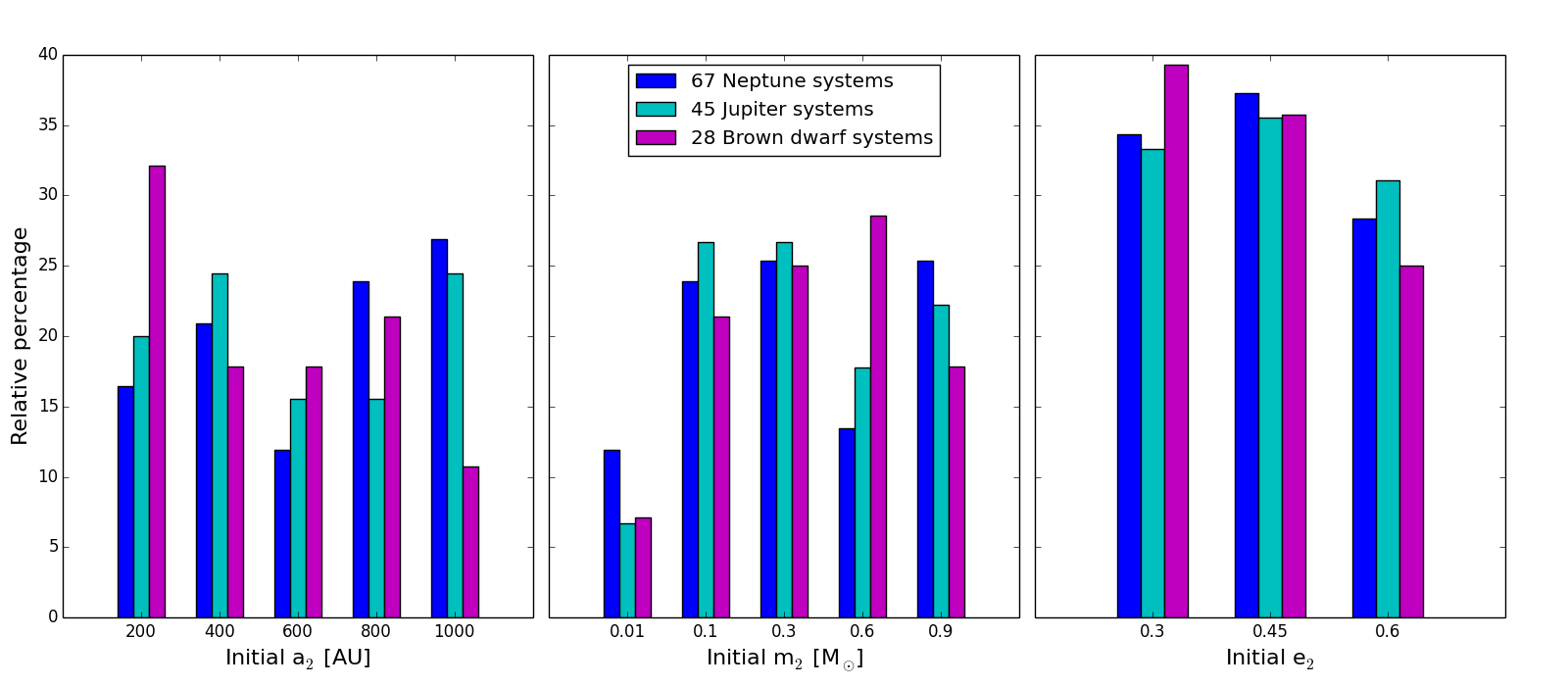}
	\caption{Percentages of Neptune, Jupiter and brown dwarf systems (relative to total of grazing systems) for the values of $m_2$, $a_2$ and $e_2$. For example, in $33\%$ out of $28$ systems with a brown dwarf (i.e., 9 systems), the location of $m_2$ is $200$ AU. }
	\label{fig:m3_hist}
\end{figure}

We now examine constraints on the mass of the third, more distant companion, as well as its eccentricity and semi-major axis. Figure \ref{fig:m3_hist} shows the percentage of grazing systems (relative to the total number of grazing brown dwarfs, Jupiters and Neptunes) which contain different values of $m_2$, $a_2$ and $e_2$. Our results indicate that most grazing Neptunes were caused by far away ($a_2$ $\sim$ 800-1000 AU), sub-solar mass companions ($m_2$ > 0.1 \msun, see middle panel of Figure \ref{fig:m3_hist}). This is expected because a less massive tertiary would be torqued by $m_1$ \citep[e.g.,][]{Naoz2016,Naoz2017AJ....154...18N}. Thus, a more massive $m_2$ is an expected consequence of the EKL mechanism. On the other hand, grazing brown dwarfs are more likely to take place for systems in which the tertiary is closer (as shown in Figure \ref{fig:m3_hist}, right panel). This is also an expected consequence of the EKL mechanism \citep[e.g.,][]{Naoz2016}. In this case, the ratio  $m_2/m_1$ is smaller than those systems with Neptune-mass inner companions, and thus we need a closer tertiary for EKL effects to induce high eccentricities.

\begin{landscape}

\begin{figure}
	\centering
	\includegraphics[width=1.2 \textwidth]{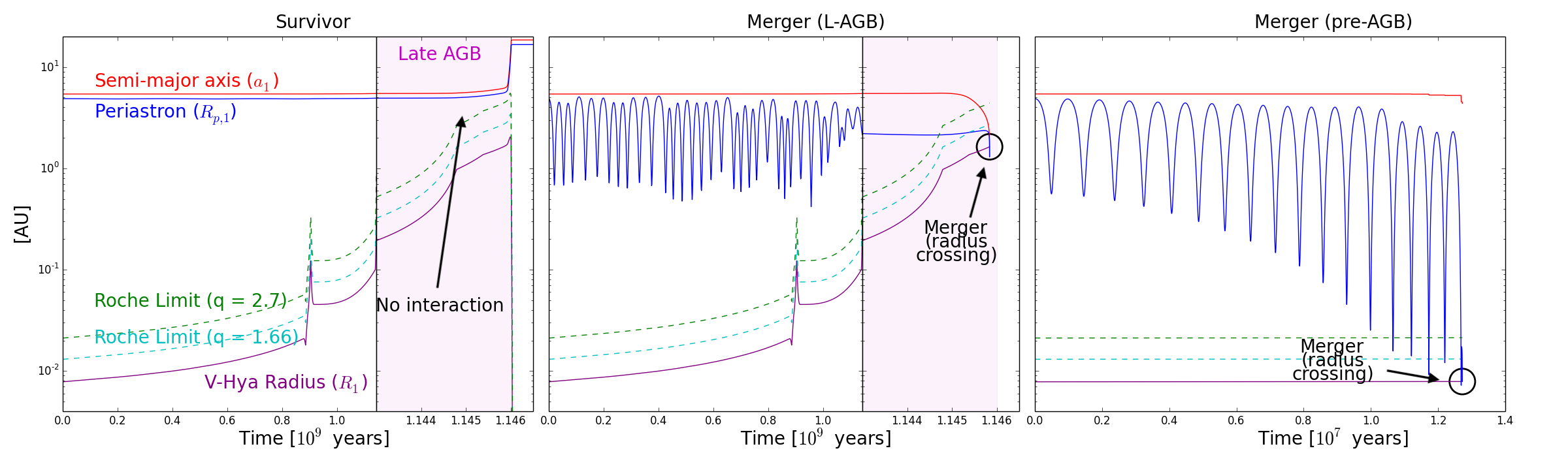}
	\caption{Example of orbital evolutions of survivor and merger systems.  The purple shaded region indicates the Late-AGB phase, which lasts for $\sim$ 3 Myrs. 
		\textit{Left}: system in which the inner companion's periastron (blue line) never crosses \hydra's Roche limit (green dashed line). The semi-major axis of the inner orbit (red line) increases substantially at $\sim$ 1.146 Gyrs. This is because at this time the star sheds most of its mass and becomes a white dwarf. Initial system parameters for this system are $m_1$ = 0.001 \msun, $m_2$ = 0.01 \msun, $a_2$ = 400 AU, $e_2$ = 0.3, $i$ = 175 degrees.
\textit{Middle}: system in which the  inner companion merges with \hydra\ during the L-AGB phase. Initial system parameters are $m_1$ = 0.001 \msun, $m_2$ = 0.01 \msun, $a_2$ = 200 AU, $e_2$ = 0.45 $i$ = 70 degrees. \textit{Right}: system in which strong EKL oscillations of the inner orbit's periastron prompts the companion to merge with \hydra\ during the Main Sequence phase. Initial system parameters are $m_1$ = 0.01 \msun, $m_2$ = 0.3 \msun, $a_2$ = 200 AU, $e_2$ = 0.6, $i$ = 70 degrees.  }
	\label{fig:sur_merge}
\end{figure}
\begin{figure}
	\centering
	\includegraphics[width=1.2 \textwidth]{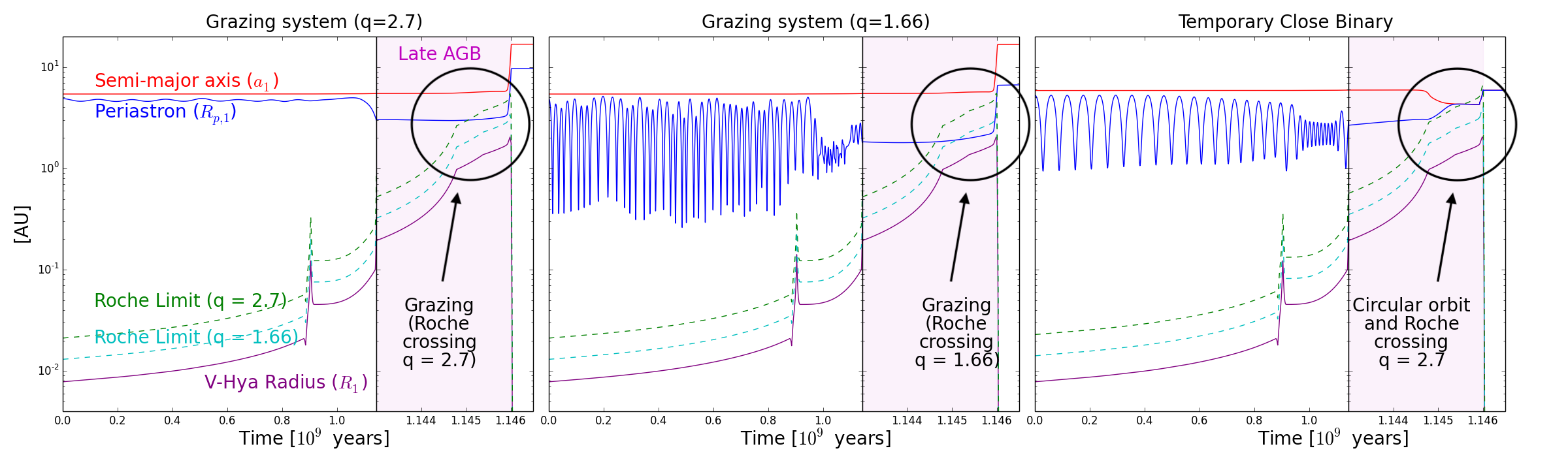}
	\caption{Example orbital evolutions of ``grazing" systems.  The purple shaded region indicates the Late-AGB phase, which lasts for $\sim$ 3 Myrs.
		\textit{Left}: system in which the inner companion's periastron (blue line) crosses \hydra's Roche limit (green dashed line, $q$ = 2.7) during the L-AGB phase. Initial system parameters are $m_1$ = 0.001 \msun, $m_2$ = 0.01 \msun, $a_2$ = 400 AU, $e_2$ = 0.6, $i$ = 105 degrees. 
		\textit{Middle}: system in which the inner companion's periastron (blue line) crosses \hydra's Roche limit (cyan dashed line, $q$ = 1.66) during the L-AGB phase. Initial system parameters are $m_1$ = 5\e{-5} \msun, $m_2$ = 0.01 \msun, $a_2$ = 200 AU, $e_2$ = 0.6, $i$ = 35 degrees. 
		\textit{Right}: system in which the inner companion's orbit circularizes, and its semi-major axis (red line) crosses the Roche limit of \hydra\ (green dashed line, $q$ = 2.7). Initial system parameters are $m_1$ = 0.6 \msun, $m_2$ = 0.6 \msun, $a_2$ = 1000 AU, $e_2$ = 0.6, $i$ = 70 degrees.   }
	\label{fig:grazing}
\end{figure}

\end{landscape}

\onecolumn
\begin{figure}
	\includegraphics[width=.51 \textwidth]{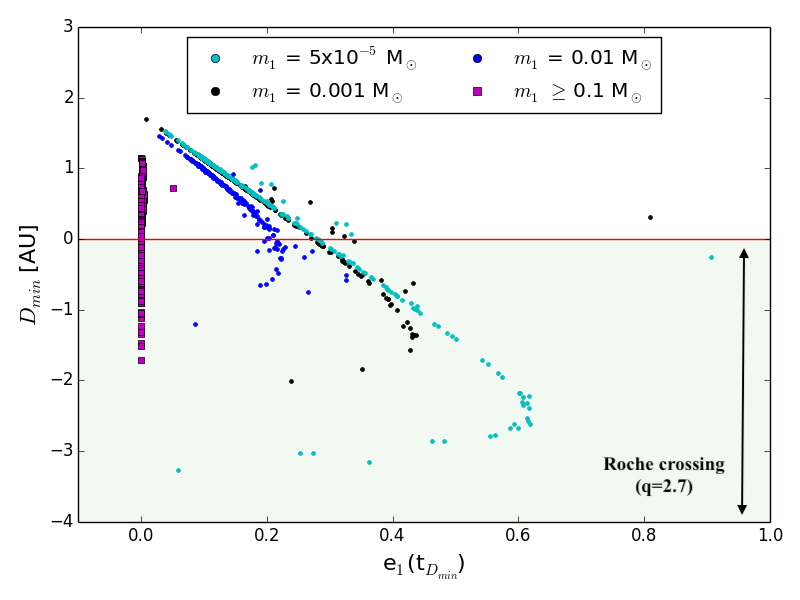}
	\includegraphics[width=.5 \textwidth]{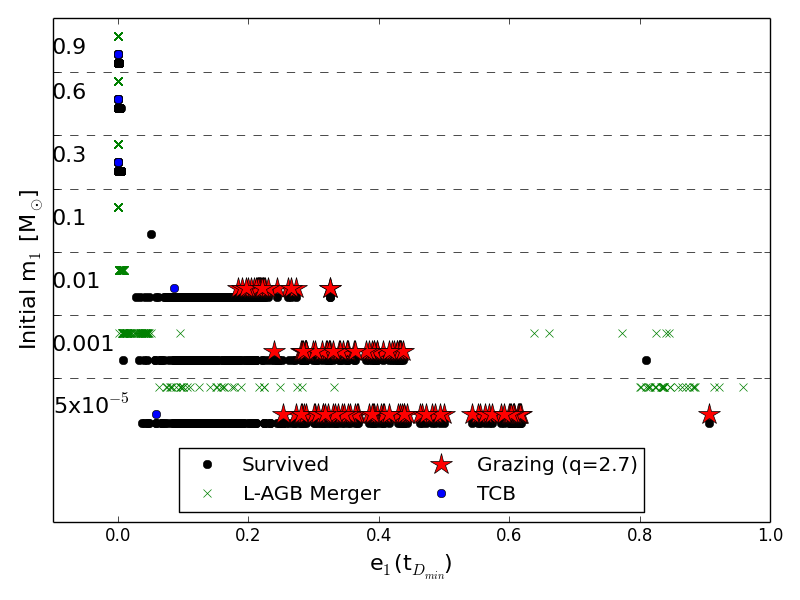}
	\caption{Illustration of how the end state of the system depends on the mass of the inner companion ($m_1$). \textit{Left}: Plot of $D_{min}$ (Equation \ref{eq:dmin}) vs the eccentricity of the inner orbit ($e_1$) at the time of minimum periastron distance to the primary's Roche limit ($t_{Dmin}$, using $q$ = 2.7). We do not include any merger system in this panel. Green shaded region represents the Roche limit crossing. 
	\textit{Right}: Plot of outcomes as a function of initial $m_1$ vs and eccentricity of the inner orbit ($e_1$) at the time of minimum periastron distance to the primary's Roche limit ($t_{Dmin}$, using $q$ = $2.7$). 
Stellar-mass ($m_1$ $>$ $0.1$ \msun, and a few planets) companions produce TCBs (blue dots), while planets (and brown dwarfs, all with $m_1$ $<$ $0.1$ \msun) produce grazing systems (red stars). Neptune-mass objects achieve the highest eccentricities. Green crosses (x) represent L-AGB mergers, and black dots ($\bullet$) represent survivor systems (like those in the left panel of Figure \ref{fig:sur_merge}).}
	\label{fig:correlations}
\end{figure}

\twocolumn

\section{Implications for the \hydra\ system} \label{sec:implications}
\subsection{Observational interpretation of the orbit}
The model introduced by \cite{Sahai2016} suggests that the observed ejections from \hydra\ are associated with the periastron passage of an unseen companion in an eccentric orbit. Furthermore, their study suggested that the eccentricity of the companion has to be relatively large, $e$ $\gtrsim$ 0.6, in order for the companion to approach the primary within its stellar envelope at periastron. 

However, here we relax the need for the companion to reach \hydra's radius, since crossing the Roche limit already provides an opportunity for interactions between  \hydra\ and the companion.
Furthermore, the low surface gravity, stellar pulsations and their associated shocks, coupled with the radiation pressure that drives \hydra's stellar wind cause an increased scale height of its atmosphere, and therefore a measurable radius is not clearly definable \citep{Zhao-Geisler2012}.

\subsection{Launching mechanism}
The launching mechanism of the observed bullet ejections is largely uncertain. 
Here we consider the consequences of a few simple proof-of-concept launching mechanisms on our proposed scenario. 
\begin{itemize}
\item 
The ballistic approximation \citep[e.g.,][]{Dosopoulou2017ApJ...844...12D} yields that the bullet speed $v_b$ should be proportional to the periastron speed $v_p$ plus the escape speed from the companion $v_{\rm esc}$, i.e,
\begin{equation}\label{eq:ballistic}
v_b \sim \sqrt{G (M_{\hydra} +m_c)\frac{1+e}{a(1-e)}} + \sqrt{\frac{2 G m_c}{r_c}} \ 
\end{equation}
where $m_c$ is the mass of the companion and $r_c$ is it's radius. A similar estimation was done by \cite{Livio1997}.
As can be seen from Equation \eqref{eq:ballistic}, the jet's velocity is highly sensitive to the  mass of the companion. In our case, each companion reaches different maximum  eccentricity (see Figure \ref{fig:correlations}), however the pericenter velocity is typically much smaller than the escape velocity. In Figure \ref{fig:ballistic} we show ejection speed according to these approximations for different companions (Neptune, Jupiter and brown dwarf).  A brown dwarf gives an ejection speed of about $230$ \kmpers, similar to the observed bullet speeds ($v \sim 200-250$ \kmpers, \citealt{Sahai2016,Scibelli2019ApJ...870..117S}).  
\begin{figure}
	\includegraphics[width=0.5 \textwidth]{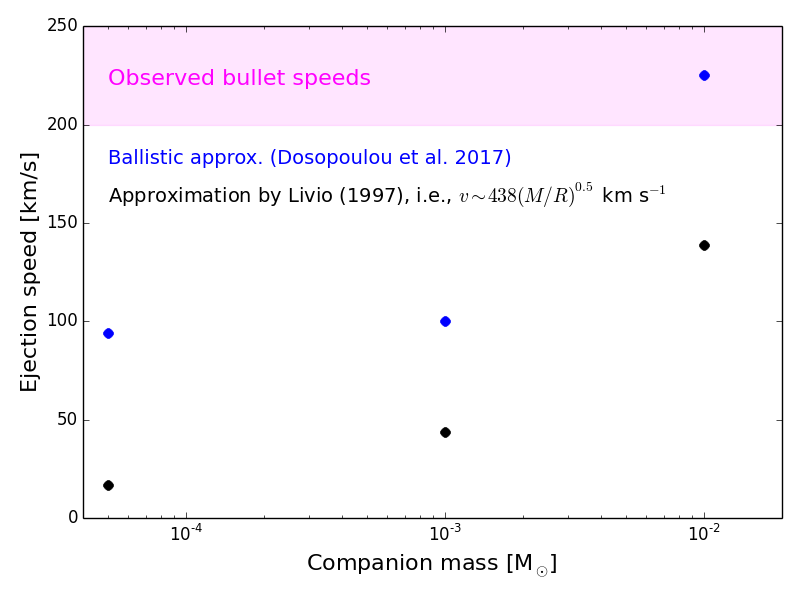}
	\caption{Jet ejection speed vs. companion mass (for Neptune, Jupiter, and brown dwarf masses). Black dots represent the approximation by \citet{Livio1997}. The blue dots represent the ballistic approximation \citep{Dosopoulou2017ApJ...844...12D}. The magenta shaded region indicates the range of the observed bullets' speeds. }
	\label{fig:ballistic}
\end{figure}
\item We also consider a magnetically driven launching mechanism. Following \cite{Fendt2003A_A...411..623F}, we consider an outflow velocity from a circum-planetary accretion disk:
\begin{dmath}\label{eq:Fendt}
v_b \approx 63 \text{ \kmpers} \left(\frac{\Phi}{5\times10^{22} \text{ G $cm^2$} } \right)^{2/3} \left(\frac{P}{4 \text{ days}} \right)^{-2/3} \times \left( \frac{\dot{M}_{out}}{10^{-3} \text{ $\dot{M}_{in}$ } } \right)^{-1/3}  \left(\frac{\dot{M}_{in}}{6\times10^{-5} \text{ $M_J$/yr}} \right) ^{-1/3}
\end{dmath}
where $\Phi$ is the magnetic flux through the accretion disk, $P$ is the outer edge disk's period, $\dot{M}_{in}$ is the inflow rate and $\dot{M}_{out}$ is the outflow rate. If the companion swings through \hydra's Roche limit and forms a transient accretion disk\footnote{Note that the timescales to form an accretion disk (t$\sim (G \rho)^{-1/2} \sim 1.2$~yr) is too long compared to the few months the companion typically spends inside the Roche limit. However, an accretion disk might be accumulated over one or more orbital revolution timescale.}, we might expect ejections due to magnetically driven jets. This mechanism is sensitive to the magnetic field of the system. For example, some studies have suggested that Jupiter's magnetic field could have been as high as $\sim$ 500 G during its early formation period \citep{Christensen2009Natur.457..167C,Batygin2018AJ....155..178B}. Furthermore, one version of the model we propose here, i.e., a Jovian planet around \hydra, might resemble the environment of a protoplanetary system, and thus, we could consider magnetic field strengths of 500 G or greater.
Similar to \cite{Fendt2003A_A...411..623F}, we adopt $10\times R_c$ as the outer disk edge. We also adopt $\dot{M}_{out}$ = $\dot{M}_{in}$, i.e., all of the accreted mass is ejected. In Figure \ref{fig:magnetic} we show the ejection speed as a function of mass accretion rate, using values of $B$ = 500 and 2000 G. As shown in the figure, accretion rates of $\sim$ $10^{-7}$ \msun/yr and a strong magnetic field ($B$= $2000$ G) are necessary for a Jovian planet to cause ejection speeds of $\sim200$ \kmpers.
\begin{figure}
	\includegraphics[width=0.5 \textwidth]{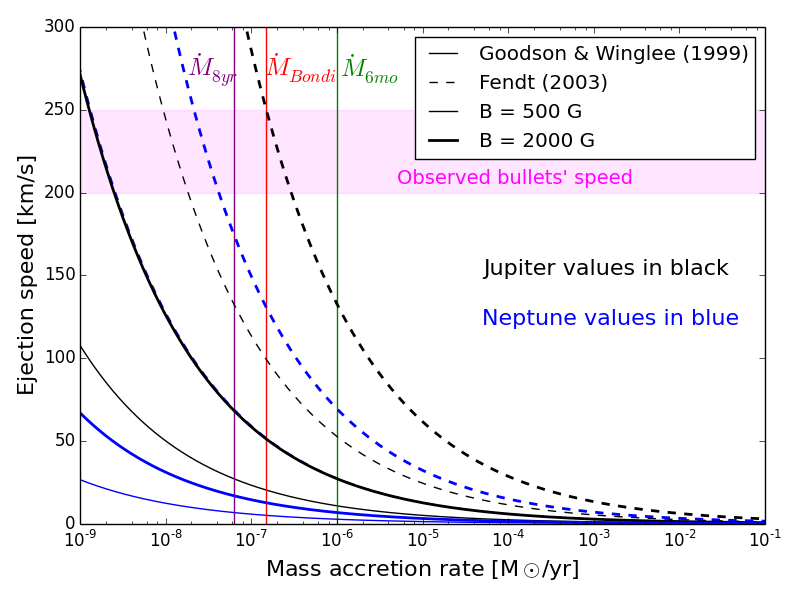}
	\caption{Plot of ejection speed vs. mass input rate. Solid lines represent ejection speed values according to Equation \eqref{eq:Goodson} \citep{Goodson1999}. Dashed lines represent values according to Equation \eqref{eq:Fendt} \citep{Fendt2003A_A...411..623F}. Black lines represent values due to a Jupiter planet, while blue represent values for a Neptune planet. Thicker lines represent values with $B$ = $2000$ $G$, while thinner lines represent values with $B$ = $500$ $G$. We also indicate some possible values of the mass accretion rate ($\dot{M}_{in}$). The vertical purple line indicates the value due to the accretion of 1 bullet mass ($\sim10^{27}$ g, \citealt{Sahai2016}) over 1 orbit ($8$ years). Similarly, the vertical green line indicates the rate of accretion of 1 bullet mass over 6 months. The vertical red line represents Bondi accretion, assuming a density $\rho$ $\sim$ $10^{-8}$ g \invcmcube (estimate taken from \citealt{Lagadec2005A&A...433..553L}). }
	\label{fig:magnetic}
\end{figure}
\item Another launching mechanism is described in \cite{Goodson1999}. In their work, they considered accretion jets from accreting magnetic young stellar objects. They found that the ejection speeds are:
\begin{dmath}\label{eq:Goodson}
	v_b = 585 P_d^{-2/3} B_{kG}^{2/3} R^{4/3} \dot{M}_8^{-1/3} \text{sin}^{4/3} \theta_{jet} \text{ km s$^{-1}$}
\end{dmath}
where $P_d$ is the stellar rotation period in days, $B_{kG}$ is the stellar surface magnetic field in kilogauss, $R$ is the stellar radius in solar radii, $\dot{M}_8$ is the jet mass flux in units of $10^{-8}$ \msun/yr, and $\theta_{jet}$ is the angle of ejection. For simplicity, we assume $\sin{\theta_{jet}}$ = $1$, and a maximally rotating Jovian planet ($P_d$ $\approx$ $0.2$ days). We consider the same values of $B$ and $\dot{M}_{in}$ as before, and show the resulting jet speed values in Figure \ref{fig:magnetic}. The speeds resulting from this mechanism are lower than those produced using Equation \eqref{eq:Fendt}.
\end{itemize}

\subsection{Relevance for companions around evolved stars}
In recent years a number of evolved stars have been observed to host Hot Jupiter companions (e.g., \citealt{Johnson2008ApJ...675..784J,Sato2008PASJ...60..539S,Johnson2010PASP..122..701J,Johnson2011AJ....141...16J,Johnson2011ApJS..197...26J,Johnson2013ApJ...763...53J,Sato2013PASJ...65...85S,Bieryla2014AJ....147...84B,Wittenmyer2015PASP..127.1021W,Wittenmyer2015ApJ...800...74W}). Recent work by \cite{Stephan2018AJ....156..128S} indicates the existence of a population of `Temporary Hot Jupiters' (THJs), of which \hydra's unseen companion could be an example. In this model, a giant planet orbiting its parent star at a significant distance is perturbed by an outer stellar companion to reach high eccentricity values. Once the parent star becomes a red giant, tidal forces bring the planet closer, becoming Hot Jupiters until they are eventually engulfed by the expanding star. However, our model here suggests that a subset of these THJs would in fact become \hydra-like systems when they begin to interact with the expanding star's extended envelope. It can, thus, be speculated that a number of observed Hot Jupiters orbiting evolved stars are indeed progenitors of future \hydra-like objects.

 \section{Summary}\label{sec:discussion}
 In this paper, we propose a dynamical configuration for the AGB star, \hydra, that accounts for its apparently periodic ``bullet" ejections. In our model, \hydra\ is part of a hierarchical triple system. The inner, $8.5$-yr period orbit is composed of \hydra\ and a low-mass companion, and they are orbited by a distant tertiary.  We have evolved a large set ($2625$) of realizations of this system, varying the masses, as well as the orbital separations, eccentricities and inclinations of the system. Our goal is to constrain the parameter space for which a \hydra-like system can occur. We include the EKL mechanism, tides, general relativity, and post-main-sequence stellar evolution. The eccentricity oscillations associated with the EKL mechanism can potentially drive the inner companion to cross inside the primary's Roche limit. Such an interaction could produce jet-like ejections via a strongly enhanced accretion episode.

Our results can be summarized as follows:
\begin{itemize}
\item \textbf{Mergers}. In $\sim37\%$ of all simulated cases, the inner companion merges with the primary, either due to strong tides during the primary's Late AGB phase, or extreme EKL effects which lead to a merger before the AGB phase. 
\item \textbf{Surviving systems}. In $\sim62\%$ of the simulated systems, the inner companion survives the evolution of the primary star until its WD phase without merging. Our investigation focuses on the subset of these cases that achieves Roche-limit crossing at periastron during the primary's Late AGB phase. 
\item \textbf{Late-AGB, grazing systems}: cases that could potentially give rise to a \hydra-like system. In these systems, the inner companion's orbit reaches a high eccentricity ($e_1>0.1$) during the primary's Late-AGB phase and crosses the primary's Roche limit during periastron passage. Thus, the inner companion could, in principle, accrete material during this period. The percentage of systems in this category depends on the properties of the evolved star's envelope as well as the details of the interactions of the close-by companion with the Roche limit \citep{Guillochon2011ApJ...732...74G,Liu2013ApJ...762...37L}. We characterize  this by adopting two different numerical pre-factors for the Roche limit. Specifically, in Eq. (\ref{eq:roche}), we find that for  $q= 2.7$ ($q=1.66$), $\sim5\%$  ($\sim1\%$) of all simulated systems may become \hydra-like systems.  It is not surprising that only small percentage of the systems present this behavior as we are constraining ourselves to a very short timescale  in the evolution of the star (the Late-AGB phase represents 1/400 of our simulation timescale). Grazing systems only occur with brown dwarfs, Jupiter and Neptune-mass inner companions (Figure \ref{fig:correlations}).  We can then estimate a possible mass and semimajor axis of the distant tertiary (Figure \ref{fig:m3_hist}). For example, we find more systems with a brown dwarf inner companion and a relatively close ($a_2\sim200$ $AU$), stellar-mass ($m_2\sim0.6$ \msun) tertiary. 
\item \textbf{Late AGB, temporary close binaries (TCBs)}. Unlike the systems in which their orbit circularizes and shrinks, in TCBs tides work to circularize the inner orbit during the primary's Late AGB phase, and the secondary migrates inside the primary's Roche limit, but a merger does not occur. Specifically, the mass loss expands the semi-major axis, which helps the companion elude engulfment. 
We find these systems only when using a value of $q$ = $2.7$ in the definition of the Roche limit, and in $\sim2\%$ of all cases we analyzed. While these cases would not produce \hydra-like systems, they are likely to end up as common envelope configurations because of the drag encountered inside the Roche limit. Most systems in this category contain stellar-mass inner companions ($m_1 > 0.1$ \msun).
\end{itemize}
 
Finally, we consider some possible launching mechanisms that could give rise to \hydra-like ejections. In particular, we examine a simple ballistic approach, as well as magnetically driven ejection processes that were suggested in the literature for proto-planetary systems and young stellar objects. Interestingly, we find (Figure \ref{fig:ballistic}) that a simple ballistic mechanism can produce the observed ejection velocity for a brown dwarf companion. However, this version of our model causes tension with the companion's suggested eccentricity ($e_1>0.6$, \citealt{Sahai2016}), since our results indicate that tidal effects may limit a brown dwarf companion's eccentricity below $\sim0.2$. Note that \cite{Sahai2016} suggested an eccentricity that was based on the companion approaching \hydra's envelope at periastron ($R_\hydra\sim2$ $AU$). We relax this condition in our investigation. 

Nevertheless, our results indicate that Jovian companions achieve the highest eccentricities during \hydra's Late AGB phase ($e_1\sim0.6$ for Neptune-mass companions, in agreement with the eccentricity suggested by \citealt{Sahai2016}). Here, a purely ballistic ejection does not produce speeds that match observations. Instead, a strong magnetic field ($B>500$ $G$) is necessary for magnetically driven outflows from a Jovian companion to achieve speeds of $\gtrsim200$ \kmpers  (Figure \ref{fig:magnetic}). This prediction can be used to distinguish competing mechanisms. 
 
This proof-of-concept study suggests that \hydra-like ejections can result from EKL-induced interactions between AGB stars and Jovian or brown dwarf companions. The model presented here also provides a framework to explain the dynamics occuring in interacting binary systems in which the companion is a stellar, sub-stellar or planetary object. This includes, for example: planets engulfed  by giant stars  \citep[e.g.,][]{Soker1984MNRAS.210..189S, Livio2002ApJ...571L.161L,Gaudi2017Natur.546..514G,Stephan2018AJ....156..128S},  the influence of planets on horizontal giant branch morphology \citep[e.g.,][]{Soker1998AJ....116.1308S}, as well as binary progenitor models of bipolar planetary nebulae \citep[e.g.,][]{Morris1987,Soker1998ApJ...496..833S}.

\section{Acknowledgments}
The authors want to thank the reviewer whose comments and suggestions helped improve and clarify this manuscript.
This work used computational and storage services associated with the Hoffman2 Shared Cluster provided by the UCLA Institute for Digital Research and Education's Research Technology Group.
This material is based upon work supported by the National Science Foundation Graduate Research Fellowship Program under Grant No. DGE-1144087.
S.N. and A.P.S. acknowledge partial support from the NSF through Grant AST-1739160.

\bibliographystyle{mnras}
\bibliography{references}

\bsp	
\label{lastpage}

\end{document}